\documentclass[11pt] {article}
\usepackage{palatino}
\usepackage{amsmath}
\usepackage{graphicx,epsfig}

\begin{document}
\title{Riemann-Liouville Fractional Einstein Field Equations}
\author{Joakim Munkhammar \footnote{Studentstaden 23:230, 752 33 Uppsala, Sweden; joakim.munkhammar@gmail.com}\\
}

\begin{small}
\date{\today}
\end{small}

\maketitle


\begin{abstract}
In this paper we establish a fractional generalization of Einstein
field equations based on the Riemann-Liouville fractional
generalization of the ordinary differential operator
$\partial_\mu$. We show some elementary properties and prove that
the field equations correspond to the regular Einstein field
equations for the fractional order $\alpha = 1$. In addition to
this we show that the field theory is inherently non-local in this
approach. We also derive the linear field equations and show that
they are a generalized version of the time fractional
diffusion-wave equation. We show that in the Newtonian limit a
fractional version of Poisson's equation for gravity arises.
Finally we conclude open problems such as the relation of the
non-locality of this theory to quantum field theories and the
possible relation to fractional mechanics.
\end{abstract}


\section{Introduction}
General relativity as a theory of gravitation is perhaps the most
successful theory in physics to date \cite{Gron}. Therefore many
generalizations of general relativity have been created in order
to possibly incorporate additional physics
\cite{Munkhammar0,Einstein1}. The various generalizations includes
for example adding extra dimensions and making the metric complex
\cite{Munkhammar0,Einstein1,Gron}.

\vspace{10pt}

The modern formulation of fractional calculus was formulated by
Riemann on the basis of Liouville's approach
\cite{Munkhammar1,Munkhammar2,WHITE}. Despite its long history it
has only recently been applied to physics where areas including
fractional mechanics and fractional quantum mechanics have been
established \cite{Laskin,Oparnica,Rabei}. Certain approaches to
fractional Friedmann equations, fractal differential geometry and
fractional variational principles have recently been developed
\cite{Baleanu,El-Nabulsi,Rabei}.

\vspace{10pt}

In this paper we shall utilize a fractional calculus operator to
generalize the differential geometry approach to general
relativity. Similar developments, in particular regarding
fractional Friedmann equations, have been made with different
kinds of fractional differentiation operators \cite{Kobelev},
however we shall utilize the simplest possible generalization
using a special form of the Riemann-Liouville fractional
derivative.


\section{Theory}
\subsection{Fractional Calculus}
Fractional calculus is the non-integer generalization of an
$N$-fold integration or differentiation \cite{Munkhammar1}. We
shall use a special case of the Riemann-Liouville fractional
differential operator as the generalization of the differential
operator $\partial_\mu$ used in general relativity. If
$f(x_\mu)\in C([a,b])$ and $-\infty <x_\mu<\infty$ then:
 \begin{equation}\label{Frac}
   \partial_\mu^{\alpha} f := \frac{1}{\Gamma (1-\alpha)} \frac{d}{dx_\mu}
   \int_{-\infty}^{x_\mu} \frac{f(t)} {(x_\mu-t)^{\alpha}} \, dt,
 \end{equation}
for $\alpha \in ]0,1[$ is here called the fractional derivative of
order $\alpha$ (It is a special case of the general
Riemann-Liouville derivative, see
\cite{Munkhammar1,Munkhammar2,WHITE} for more information). For
$\alpha = 1$ we define $\partial^\alpha_\mu = \partial_\mu$ and we
have the ordinary derivative.

\subsection{Fractional Einstein field equations}
We shall assume spacetime to be constituted by a metric of the
same type as for ordinary Einstein field equations \cite{Gron}:
\begin{equation}\label{Metric}
ds^2 = g_{\mu \nu} dx^\mu dx^\nu,
\end{equation}
with the usual condition for the covariant metric
\cite{Munkhammar0,Gron}:
\begin{equation}
g_{\mu \nu} g^{\mu \rho} = \delta^\rho_{\nu}.
\end{equation}
The metric can also be characterized via vielbein according to:
\begin{equation}
g_{\mu \nu} = e^a_\mu e^b_\nu \eta_{ab}.
\end{equation}
The dynamics is governed by the Einstein field equations (here
with fractional parameter $\alpha$):
\begin{equation}\label{EFE}
R_{\mu \nu}(\alpha) - \frac12 g_{\mu \nu} R (\alpha) = \frac{8 \pi
G}{c^4} T_{\mu \nu} (\alpha).
\end{equation}
In this fractional approach we shall assume that the Riemann
tensor is constructed from the metric on the basis of the
Riemann-Liouville fractional derivatives \eqref{Frac}. We have the
Riemann tensor as follows:
\begin{equation}
R_{\mu \nu}(\alpha) = R^\lambda_{\mu \lambda \nu}(\alpha) =
\partial_\lambda \Gamma^\lambda_{\mu \nu}(\alpha) - \partial_\nu
\Gamma^\lambda_{\mu \lambda}(\alpha) + \Gamma^{\lambda}_{\sigma
\lambda} (\alpha) \Gamma^{\sigma}_{\mu \nu}(\alpha) -
\Gamma^\lambda_{\sigma \nu} (\alpha) \Gamma^{\sigma}_{\lambda
\mu}(\alpha),
\end{equation}
where we define the Christoffel symbol as:
\begin{equation}\label{Christoffel}
\Gamma^{\mu}_{\nu \lambda}(\alpha) = \frac12 g^{\mu \rho}
(\partial^\alpha_\nu g_{\rho\lambda} +
\partial^{\alpha}_{\lambda} g_{\rho\nu} - \partial^\alpha_\rho
g_{\nu \lambda}),
\end{equation}
and $\partial^\alpha_\mu$ is the fractional derivative
\eqref{Frac}. We may also construct a covariant fractional
derivative:
\begin{equation}
\nabla^\alpha_\nu A_\mu = \partial^\alpha_\nu A_\mu - A_\beta
\Gamma^\beta_{\mu \nu} (\alpha),
\end{equation}
for some four vector $A_\mu$. We also have the fractional geodesic
equation:
\begin{equation}\label{Geodesic}
\frac{d^2 x^\alpha}{d \tau^2} + \Gamma^{\alpha}_{\beta \sigma}
(\gamma) \frac{dx^\beta}{d \tau}\frac{dx^\sigma}{d \tau} = 0.
\end{equation}
The correspondence to the traditional Einstein field equations is
when $\partial^\alpha_\mu$ is replaced by the traditional
derivative $\partial_\mu$ which is equivalent to $\alpha = 1$ by
definition. A very special feature of this field theory is that it
is non-local since the fractional derivative \eqref{Frac} utilizes
information from the metric at great distances to create the
fractional derivative.


\subsection{Linearized field equations}
We can write the metric \eqref{Metric} as:
\begin{equation}
g_{\mu \nu} = \eta_{\mu \nu} + h_{\mu\nu},
\end{equation}
where $\eta_{\mu \nu}$ is the Minkowski metric and $h_{\mu \nu}$
is higher order terms. To first order in $h_{\mu \nu}$ we may
neglect products of Christoffel symbols and thus arrive at
\cite{Gron}:
\begin{equation}
R_{\alpha \mu \beta \nu}(\gamma) = \partial^\gamma_\beta
\Gamma_{\alpha \mu \nu}(\gamma) - \partial^\gamma_\nu
\Gamma_{\alpha \mu \beta} (\gamma)
\end{equation}
where the Christoffel symbol $\Gamma_{\alpha \mu \nu} (\gamma)$ is
fractional of order $\gamma$ according \eqref{Christoffel} (here
linearized):
\begin{equation}
\Gamma_{\alpha \mu \nu}(\gamma) = \frac12 (\partial^\gamma_\nu
h_{\mu \alpha} + \partial^\gamma_\mu h_{\nu \alpha} -
\partial^\gamma_\alpha h_{\mu \nu})
\end{equation}
This brings the the Riemann tensor to:
\begin{equation}
R_{\alpha \mu \beta \nu}(\gamma) = \frac12(\partial^\gamma_\mu
\partial^\gamma_\beta h_{\nu \alpha} + \partial^\gamma_\alpha
\partial^\gamma_\nu h_{\mu \beta} - \partial^\gamma_\alpha
\partial^\gamma_\beta h_{\mu \nu} - \partial^\gamma_\mu
\partial^\gamma_\nu h_{\alpha \beta}).
\end{equation}
We also have the Ricci tensor (in this approximation):
\begin{equation}
R_{\mu \nu} (\gamma) = \frac12 (\partial^\gamma_\alpha
\partial^\gamma_\mu h^{\alpha}_\nu +´\partial^\gamma_\alpha
\partial^\gamma_\nu h^\alpha_\mu - \partial^\gamma_\mu
\partial^\gamma_\nu h - \eta^{\alpha \beta} \partial^\gamma_\alpha \partial^\gamma_\beta h_{\mu \nu})
\end{equation}
where $h \equiv h^\alpha_\alpha$. We also have the Ricci scalar
as:
\begin{equation}
R = \partial^\gamma_\mu \partial^\gamma_\nu h^{\mu \nu}
-\eta^{\alpha \beta} \partial^\gamma_\alpha \partial^\gamma_\beta
h_{\mu \nu}.
\end{equation}
The field equations \eqref{EFE} now appears as:
\begin{equation}\label{LinEFE}
\partial^\gamma_\alpha
\partial^\gamma_\mu h^{\alpha}_\nu +´\partial^\gamma_\alpha
\partial^\gamma_\nu h^\alpha_\mu - \partial^\gamma_\mu
\partial^\gamma_\nu h - \eta^{\alpha \beta} \partial^\gamma_\alpha \partial^\gamma_\beta h_{\mu
\nu}-\eta_{\mu \nu} (\partial^\gamma_\alpha \partial^\gamma_\beta
h^{\alpha \beta} -\eta^{\alpha \beta} \partial^\gamma_\alpha
\partial^\gamma_\beta h_{\mu \nu}) = \frac{8 \pi
G}{c^4} T_{\mu \nu}(\gamma).
\end{equation}
It is useful to introduce the following:
\begin{equation}
\overline{h} = h_{\mu \nu} - \frac12 \eta_{\mu \nu} h,
\end{equation}
and if we now impose a form of fractional Lorenz gauge condition:
\begin{equation}
\partial^\gamma_\beta \overline{h}^\beta_{\alpha} = 0,
\end{equation}
the linearized field equations \eqref{LinEFE} turns out to be:
\begin{equation}
\partial^\gamma_\alpha
\partial^\gamma_\beta h_{\mu \nu} = \frac{8 \pi
G}{c^4} T_{\mu \nu}(\gamma).
\end{equation}
One might define these set of equations as the fractional
gravitoelectromagnetic equations, which is a fractional analogue
of the ordinary gravitoelectromagnetic equations in general
relativity \cite{Mashhoon}. In the situation where $T_{\mu \nu}$
vanishes we have:
\begin{equation}\label{FracWave}
\partial^\gamma_\alpha
\partial^\gamma_\beta h_{\mu \nu} = 0,
\end{equation}
which is a generalization of the time fractional diffusion-wave
equation (see \cite{Mainardi}). If we let $\gamma = 1$ in
\eqref{FracWave} then we get:
\begin{equation}
\Big( \nabla^2-\frac{\partial^2}{\partial t^2} \Big) h_{\mu \nu} =
0,
\end{equation}
which is the regular wave equation for gravitational waves in
General Relativity \cite{Gron}.
%


\subsection{Newtonian approximation}
The field equations \eqref{EFE} can be written in a trace-reversed
form as:
\begin{equation}\label{TraceEFE}
R_{\mu \nu} (\gamma) = \frac{8\pi G}{c^4}\Big( T_{\mu \nu}(\gamma)
- \frac12 T (\gamma) g_{\mu \nu} \Big).
\end{equation}
In order to find the Newtonian limit we let a test particle
velocity be approximately zero:
\begin{equation}
\frac{d x^\beta}{d \tau} \approx \Big(\frac{dt}{d\tau},0,0,0\big),
\end{equation}
and thus:
\begin{equation}
\frac{d}{dt} \frac{dt}{d \tau} \approx 0.
\end{equation}
Furthermore we shall assume that the metric and its fractional
derivatives are approximately static and that the square
deviations from the Minkowski metric are negligible. This gives us
the fractional geodesic equation as:
\begin{equation}
\frac{d^2 x^i}{d t^2} \approx - \Gamma^i_{00} (\alpha),
\end{equation}
and the fractional derivative of the Newtonian potential $\phi$ is
then:
\begin{equation}
\partial^\gamma_i \phi \approx \frac12 g^{i \alpha} (g_{\alpha 0,0} +
g_{0\alpha,0} - g_{00,\alpha}) \approx -\frac12 g^{ij}g_{00,j}
\approx \frac12 g_{00,i}.
\end{equation}
We may also assume that:
\begin{equation}
g_{00} \approx -c^2 - 2\phi,
\end{equation}
holds. We only need the $00$-component of the trace-reversed field
equations \eqref{TraceEFE}:
\begin{equation}
R_{00} (\gamma) = \frac{8 \pi G}{c^4} \Big(T_{00} (\gamma) -
\frac12 T(\gamma)g_{00} \Big).
\end{equation}
The stress-energy tensor in a low-speed and static field
approximation becomes:
\begin{equation}
T_{00}(\gamma) \approx \rho(\gamma) c^2,
\end{equation}
where $\rho(\alpha)$ is a form of fractional generalization of
matter density $\rho$ (also note that for $\alpha = 1$ the
ordinary matter density $\rho$ is obtained). We thus get the
$00$-component of the Riemann tensor as:
\begin{equation}
\frac{8 \pi G}{c^4}\Big(T_{00}(\gamma) - \frac12T(\gamma) g_{00}
\Big) \approx 4 \pi G \rho (\gamma).
\end{equation}
We also have:
\begin{equation}
\partial^\gamma_i \partial^\gamma_i \phi \approx \Gamma^i_{00,i}
\approx R_{00}(\gamma) \approx 4 \pi G \rho(\gamma),
\end{equation}
which in the case $\gamma = 1$ turns out to be the traditional
Poisson's equation for gravity:
\begin{equation}
\partial_i \partial_i \phi = \nabla^2 \phi = 4 \pi G \rho,
\end{equation}
which is the expected result.

\section{Conclusions}
We have established a fractional generalization of Einstein field
equations based on the Riemann-Liouville fractional generalization
of the differential operator $\partial_\mu$. We showed some
special properties including the correspondence to conventional
Einstein field equations for $\alpha = 1$. We showed that the
field equations by necessity are non-local due to the non-local
nature of the Riemann-Liouville fractional derivative. In fact the
field is non-local even for $\alpha = 1-\epsilon$ for some
arbitrarily small $\epsilon$. In this situation the field
equations are practically equivalent to the ordinary Einstein
field equations, but they still have the non-local property.
Furthermore we linearized the field equations and obtained a
fractional wave equation with source, which is then the fractional
equivalent of gravitoelectromagnetism in ordinary general
relativity. We also investigated the Newtonian limit and arrived
at a fractional generalization of the classical Poisson's equation
for gravity. Many open questions remain, such as if the
non-locality in some way may be connected to quantum field
theories. Another issue is if there is any connection between this
approach and the fractional stability of Friedmann equations in
ordinary Einstein field equations \cite{El-Nabulsi}. Also further
connections to fractional mechanics and other fractional and
fractal approaches to gravity are open issues.

\end{document}